# Prediction of Novel CXCR7 Inhibitors Using QSAR Modeling and Validation via Molecular Docking

Belaguppa Manjunath Ashwin Desai[a], Merla Sudha[b], Suvarna Ghosh[b] and Pronama Biswas[b]*

[a]*School of Engineering*, Dayananda Sagar University, Bengaluru, India
[b]*School of Basic and Applied Sciences*, Dayananda Sagar University, Bengaluru, India

*pronama-sbas@dsu.edu.in

*Abstract*— CXCR7, a G-protein-coupled chemokine receptor, has recently emerged as a key player in cancer progression, particularly in driving angiogenesis and metastasis. Despite its significance, currently, few effective inhibitors exist for targeting this receptor. In this study aimed to address this gap by developing a QSAR model to predict potential CXCR7 inhibitors, followed by validation through molecular docking. Using the Extra Trees classifier for QSAR modeling and employing a combination of physicochemical descriptors and molecular fingerprints, compounds were classified as active or inactive with a high accuracy of 0.85. The model could efficiently screen a large dataset, identifying several promising CXCR7 inhibitors. The predicted inhibitors were further validated through molecular docking studies, revealing strong binding affinities, with the best docking score of -12.24 ± 0.49 kcal/mol. Visualization of the docked structures in both 2D and 3D confirmed the interactions between the inhibitors and the CXCR7 receptor, reinforcing their potential efficacy.

## I. Introduction

According to the Global Cancer Observatory (GLOBOCAN) 2022, cancer is responsible for approximately 9.7 million deaths annually worldwide. With a projected 35 million new cases by 2050, discovering new drugs to inhibit cancer progression is crucial [1]. Chemokines are small signaling proteins that directly affect tumor growth and metastasis [2]. One such chemokine, C-X-C motif chemokine ligand 12 (CXCL12), is primarily responsible for formation of blood vessels and spread of cancer cells of cancer cells from one place to another. Previously, CXCL12 was thought to bind only to the receptor CXCR4, but recent studies have identified an additional receptor, CXCR7, which shows significantly better binding affinity to CXCL12 [3].

CXCR7 is a unique chemokine-specific receptor that is overexpressed in multiple cancers such as breast and lung cancer, pancreatic cancer, glioma, thyroid cancer, etc., [3,4]. CXCR7 upregulates AKT and EGFR signaling pathways, increasing cell proliferation and survival. Moreover, CXCL12 regulates the expression of interleukin-8 (IL-8) and vascular endothelial growth factor (VEGF), contributing to formation of blood vessels. Additionally, its role in cell adhesion and migration suggests that inhibiting CXCR7 holds significant potential for preventing cancer metastasis [5]. However, currently, no commercially approved drugs are available in the

To address this urgent need for CXCR7 inhibitors, Quantitative Structure-Activity Relationship (QSAR) modeling was employed, a commonly used computational method in drug discovery [6]. Using QSAR, a classification model based on molecular fingerprints and physicochemical descriptors was developed. After refining the model and achieving high classification accuracy, ten lead compounds were found with promising inhibitory activity. After identifying promising candidates, molecular docking studies were performed to further elucidate the interactions between the predicted inhibitors and the CXCR7 receptor. This combined approach leverages both computational modeling and docking techniques to explore new avenues for targeting CXCR7 in cancer therapy.

## II. Materials and Methods

### A. Data Collection and Cleaning

The ChEMBL database was utilized to identify potential inhibitors and to download the data on this target and their inhibitory activities (https://www.ebi.ac.uk/chembl/). The target with ChemBL-ID "CHEMBL2010631" was chosen for its largest IC50 data from Homo sapiens which covered a wide range of compounds and assay conditions. This data was filtered particularly for IC50 values and then compiled into a DataFrame and saved as "Rawdata_CXCR7.csv", which consisted of 249 molecules. Then, the dataset was cleaned by removing entries with missing IC50 values, invalid measurements (e.g., "<" or ">"), and entries without SMILES. IC50 values were standardized to nanomolar (nM). Redundant compounds were evaluated for consistency, and compounds with high IC50 standard deviations were excluded. The final dataset of 155 entries was saved as "Cleaned_CXCR7.csv".

### B. Descriptors and Fingerprint Calculation

Using RDKit, 211 molecular descriptors were calculated from SMILES strings (https://www.rdkit.org/). These descriptors represented key physicochemical and structural properties. Following this, invalid molecules were tracked and excluded. The final set of descriptors was saved as "RDkit_Descriptors_CXCR7.csv". Similarly, several molecular fingerprints (Morgan, Avalon, MACCS keys, topological torsion, and atom pair fingerprints) were computed to evaluate structural features. These fingerprints were concatenated and saved as "RDkit_FPs_CXCR7.csv" for subsequent modeling. Together, the descriptors and the fingerprints presented a total of 2937 features for QSAR modeling.

## C. Data Preprocessing, Feature Selection and Scaling

IC50 values were converted to pIC50 and binarized to classify compounds as active (pIC50 ≥ 6) or inactive (pIC50 < 6). All columns were checked for any missing or infinite values, and only the columns with finite IC50 values were considered. The final dataset contained 63 active (Class 1) and 92 inactive (Class 0) compounds. A variance threshold was applied to remove low-variance features, followed by mutual information-based filtering to select relevant features. After scaling, 331 features remained for modeling. The preprocessed data was saved as "Classification_data_CXCR7.csv".

## D. Model Training and Testing

Various machine learning models (Logistic Regression, Random Forest, Extra Trees, etc.) were trained using stratified train-test splitting (70:30) and balanced class weights to address the imbalance in the dataset. Extra Trees Classifier demonstrated superior performance and was further optimized using GridSearchCV for hyperparameter tuning. To ensure the model's generalizability, 5-fold cross-validation was applied, and the model's performance was evaluated using various metrics, including accuracy, precision, recall, and F1-score. Additionally, the Area Under the Curve (AUC) from the Receiver Operating Characteristic (ROC) curve was calculated for each fold.

## E. Tools and Libraries

This analysis was conducted using Python, leveraging RDKit for descriptor and fingerprint calculation, Scikit-learn for model building (https://scikit-learn.org), and Matplotlib for visualizations (https://matplotlib.org/). The computational analyses were executed on a high-performance server equipped with an AMD EPYC 7742 64-core processor, 503 GiB of RAM, and NVIDIA A100 GPUs operating on Ubuntu 22.04.4. All processes were carried out in a dedicated Conda environment to ensure consistency and reproducibility of the results.

## F. Dataset for Prediction

To identify novel CXCR7 inhibitors, a dataset consisting exclusively of Genuine Natural (GN) and Naturally Sourced (NS) compounds was downloaded from the MolPort database (https://www.molport.com). GN compounds occur naturally, although they may be produced synthetically, while NS compounds are isolated directly from natural sources. This dataset was subsequently tested using the trained Extra Trees Classifier model. The objective was to predict the inhibitory activity of these compounds against CXCR7 using the trained model. This allowed for the identification of potential novel inhibitors with significant biological relevance. Further, since these compounds are GN or NS, their toxicity is expected to lower, or they are expected to be biocompatible.

## G. Molecular Docking Methods

The top 10 predicted inhibitors, showing the highest probabilities of being active, were chosen for further validation through molecular docking protocol [7]. These molecules were docked against the CXCR7 target, and their binding affinities were compared to that of a known inhibitor, CCX777 (CID: 5366167) [8]. The protein structure of CXCR7 was retrieved as a Protein Data Bank (PDB) file, which was obtained from AlphaFold when searched with the UniProt ID "P25106". The protein quality was tested using PROCHECK (https://saves.mbi.ucla.edu/) and VoroMQA (https://bioinformatics.lt/wtsam/voromqa) websites. The stereochemistry of the SMILES was corrected and converted to SDF files, followed by conversion to PDBQT files with Gasteiger charges added using a command from Open Babel. CXCR7 and CCX77 were prepared using AutoDockTools-1.5.7 (https://autodock.scripps.edu/) and saved in PDBQT format. Molecular docking of CCX777 and CXCR7 was performed using a Vina Script-based method in a Conda environment using a server, which was accessed through Visual Studio Code. The molecular docking of the predicted molecules was performed in a batch using the same script-based method. Molecular docking was performed in triplicates, and the average and standard deviation of the obtained binding affinity values were calculated. The protein-ligand docked complex was analyzed to visualize amino acids and binding pockets using UCSF Chimera 1.17 (https://www.cgl.ucsf.edu/chimera/) and Discovery Studio 2021 Client (https://www.3ds.com/products/biovia/discovery-studio/visualization) (Fig. 1.).

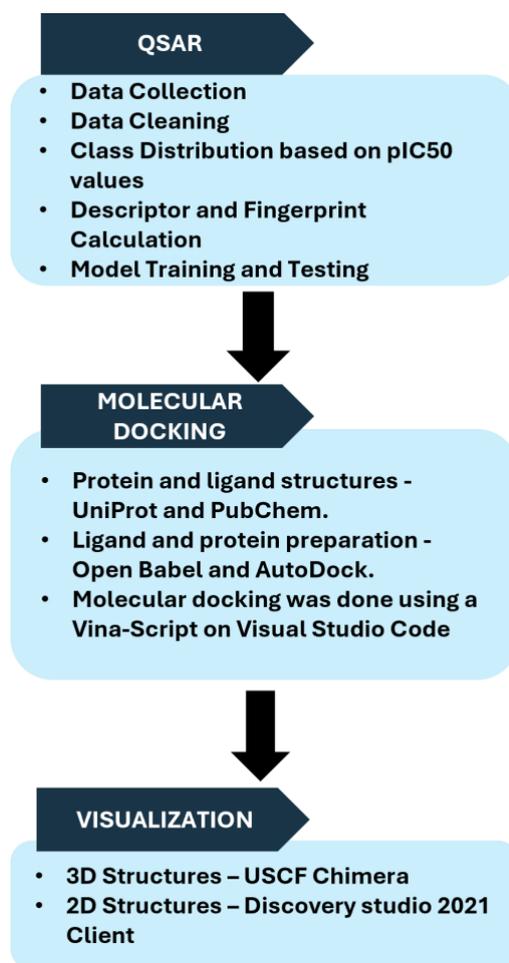

Fig. 1. Schematic representation of the QSAR, molecular docking and visualization methodology involved in the study. The figure was made using Microsoft PowerPoint® Version 2409 (Build 18025.20030).

## III. RESULTS AND DISCUSSIONS

### A. Dataset Characteristics

After the data cleaning and preprocessing steps, a total of 155 molecules were left in the dataset, among which 92 were inactive molecules (pIC50 < 6), and 63 were active (pIC50 ≥ 6). Fig. 2. illustrates the distribution of pIC50 values across both active and inactive molecules. As depicted, the majority of inactive compounds are concentrated around a pIC50 value of 5, while active compounds are more evenly distributed, particularly between pIC50 values of 6 and 9. This distribution highlights a class imbalance, which was addressed using appropriate strategies like adjusting class weights during the machine learning process.

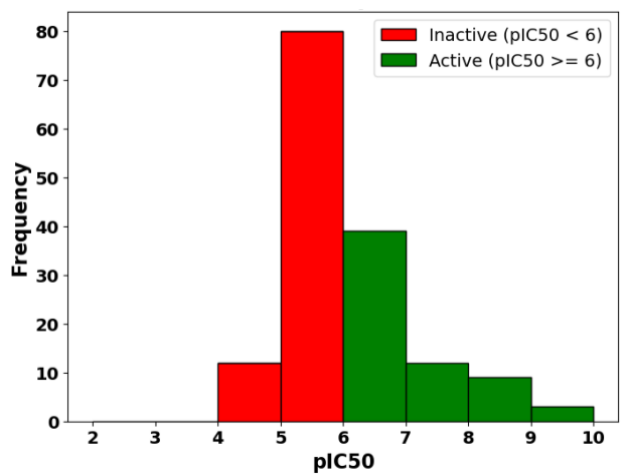

Fig. 2. Distribution of pIC50 values for active (pIC50 ≥ 6) and inactive (pIC50 < 6) molecules in the dataset after data cleaning. The dataset contains 92 inactive and 63 active molecules, with a clear concentration of inactive compounds around pIC50 = 5 and active compounds between pIC50 = 6 and 9.

### B. Performance of the Model

The Extra Trees Classifier, optimized with specifically tuned hyperparameters—{'bootstrap': False, 'max_depth': None, 'max_features': 'log2', 'min_samples_leaf': 1, 'min_samples_split': 10, 'n_estimators': 300}—demonstrated strong performance in predicting CXCR7 inhibitors across multiple cross-validation folds. The model achieved an average accuracy of 0.85 and a good ROC AUC of 0.93, as outlined in Table I. It consistently exhibited high precision (0.84 for active compounds) and recall (0.79 for active compounds), critical metrics for correctly identifying true CXCR7 inhibitors while minimizing false negatives. These high scores reflect the model's ability to accurately predict active inhibitors. The average F1-score of 0.88 for inactive compounds and 0.81 for active compounds indicates a balanced trade-off between precision and recall. This balance is particularly important in pharmacological research, where efficiently distinguishing between active and inactive compounds can significantly influence the success of drug development efforts. ROC curves from each validation fold (Fig. 3.) consistently demonstrated the classifier's strong discriminatory power, with the mean ROC curve presenting strong performance. This suggests that the Extra Trees Classifier not only accurately predicts CXCR7 inhibition but also maintains reliability across different thresholds, making it highly effective for both screening purposes and detailed pharmacological assessments.

TABLE I. SUMMARY OF AVERAGE METRICS FOR THE EXTRA TREES CLASSIFIER FOR FIVE FOLDS

| Metric | Value |
| --- | --- |
| Average Accuracy | 0.85 |
| Average Precision (Class 0) | 0.87 |
| Average Recall (Class 0) | 0.90 |
| Average F1-Score (Class 0) | 0.88 |
| Average Precision (Class 1) | 0.84 |
| Average Recall (Class 1) | 0.79 |
| Average F1-Score (Class 1) | 0.81 |
| Average ROC AUC | 0.93 |

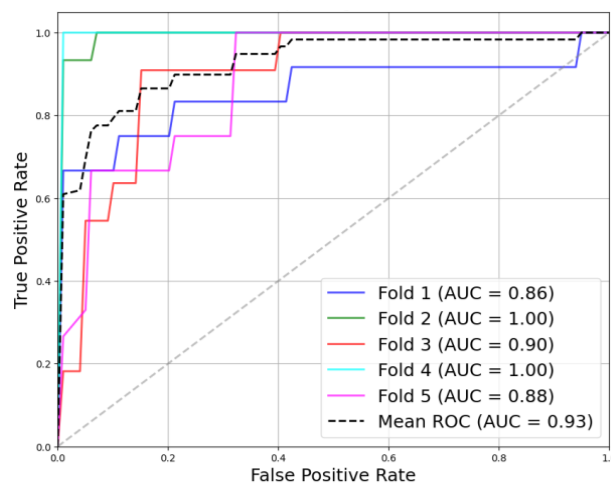

Fig. 3. Receiver Operating Characteristic (ROC) Curves for Each Fold in Cross-Validation. The curves demonstrate the model's discriminative performance across the folds with Area Under the Curve (AUC) values ranging from 0.86 to 1.

### C. Docking of Predicted Molecules

Out of the 10,290 inhibitors tested, 660 inhibitors were predicted as active using the trained model. These inhibitors were subsequently subjected to PAINS (Pan Assay Interference Compounds) filtering, which reduced the dataset to 659 molecules by removing any false positives or compounds likely to interfere in biological assays [9]. The binding affinity of the known inhibitor was observed to be -8.82 ± 0.16, which is lower compared to most of the binding affinity values obtained from docking the top 10 predicted inhibitors against CXCR7 (Table II). Binding affinity measures the strength of interaction between a ligand and a protein. It is quantitatively represented by the Gibbs free energy of binding (ΔG), where more negative values indicate stronger binding interactions [10].

From these binding affinity results, it was inferred that the predicted inhibitors had a greater inhibitory effect, reflecting better efficacy than CCX777, as they exhibited better binding affinities compared to CCX777. For example, Molecule 1 (MolPort-007-980-806) and Molecule 5 (MolPort-020-005-850) exhibited high binding affinity values of -12.24 ± 0.49 kcal/mol and -9.50 ± 0.13 kcal/mol, respectively. The minimal standard deviation demonstrated the reproducibility of the docking protocol. From the 3D visualization, it was seen that the binding pockets of CCX777 and the predicted inhibitors are the same (Fig. 4(a), (c), (e)). This indicated that both the CCX777 and the predicted inhibitors bound to the active site of CXCR7. Additionally, in the 2D visualization, it was observed that the docked complexes of CCX777, molecule 1, and molecule 5 shared common binding residues: SER 103, ASN 108, HIS 121, TYR 195, CYS 196, HIS 298, and GLN 30, as highlighted in red circles (Fig. 4.). This further supports that the predicted inhibitors bind at the same site as the known inhibitor.

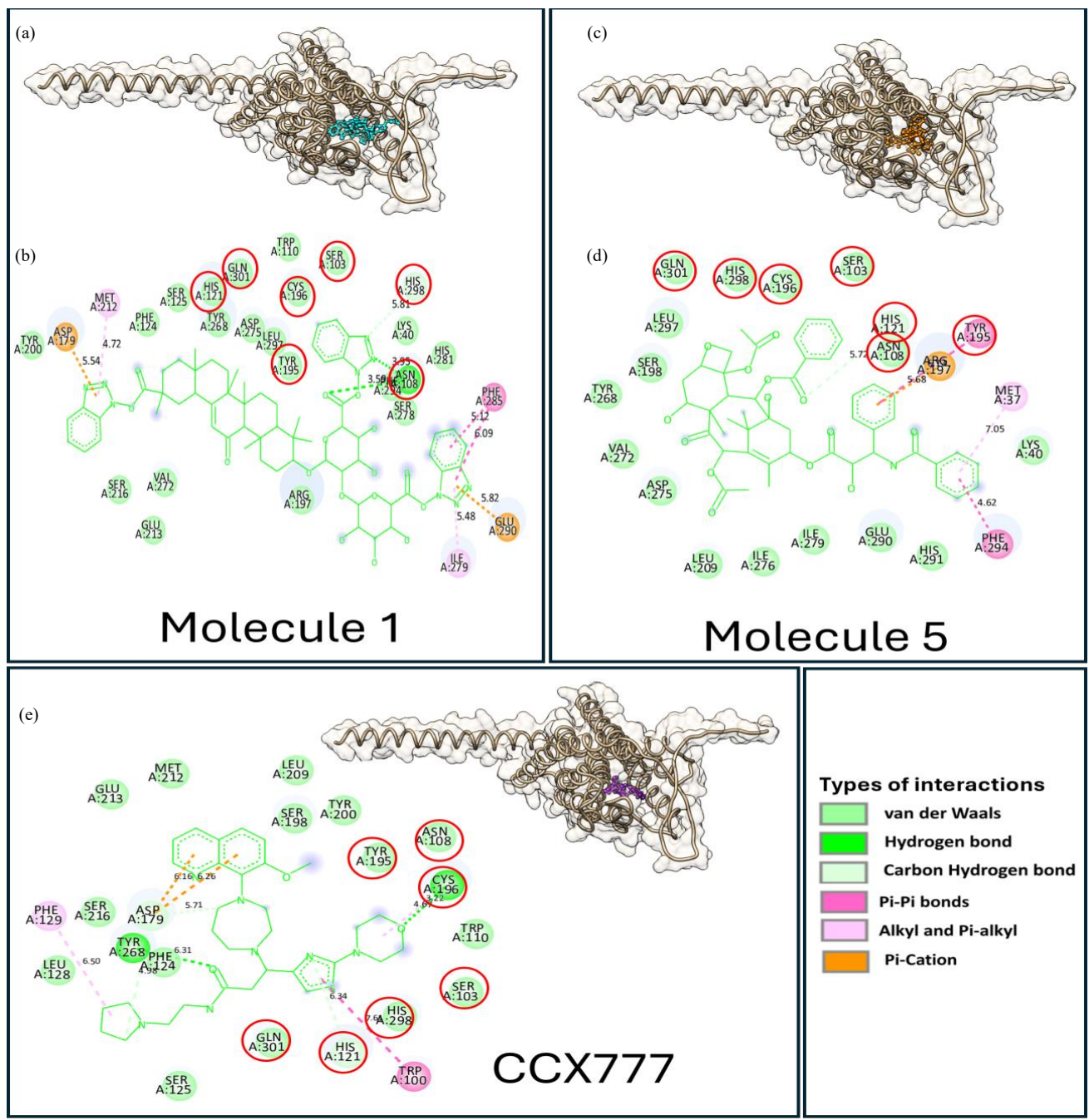

Fig. 4. Visual representation of binding pockets and binding amino acids of the docked complex. (a) 3D-visualisation of molecule 1 (MolPort-007-980-806), (b) 2D visualisation of molecule 1 (MolPort-007-980-806), (c) 3D-visualisation of molecule 5 (MolPort-020-005-850), (d) 2D visualisation of molecule 5 (MolPort-020-005-850), (e) 3D- and 2D-visualisation of CCX777.

TABLE II. BINDING AFFINITY VALUES OF CXCR7 WITH TOP 10 PREDICTED INHIBITORS

| Sl.no. | Prediction Probability | MOLPORT ID | Binding affinity (kcal/mol) |
|---|---|---|---|
| 1 | 0.359 | MolPort-007-980-806 | -12.24 ± 0.49 |
| 2 | 0.383 | MolPort-007-980-770 | -10.50 ± 0.70 |
| 3 | 0.384 | MolPort-005-945-183 | -8.20 ± 1.08 |
| 4 | 0.390 | MolPort-001-742-627 | -9.80 ± 0.02 |
| 5 | 0.390 | MolPort-020-005-850 | -9.50 ± 0.13 |
| 6 | 0.393 | MolPort-008-348-433 | -9.41 ± 0.25 |
| 7 | 0.394 | MolPort-005-945-484 | -9.98 ± 0.20 |
| 8 | 0.396 | MolPort-035-705-503 | -9.73 ± 0.51 |
| 9 | 0.396 | MolPort-039-338-170 | -8.74 ± 0.34 |
| 10 | 0.397 | MolPort-007-980-774 | -9.00 ± 0.06 |

## IV. CONCLUSIONS

Cancer remains a major global health challenge as one of the leading causes of mortality, highlighting the urgent need for innovative and effective treatment strategies. Developing inhibitors through wet lab studies is both time-consuming and expensive, necessitating faster and more efficient approaches. CXCR7 is a chemokine receptor involved in cancer progression. It is overexpressed in various cancers and contributes to cell growth, survival, migration, and blood vessel formation. Despite its significant role in cancer progression, there are no commercially available drugs targeting CXCR7. This study demonstrated the successful use of QSAR modeling and molecular docking to propose ten potential CXCR7 inhibitors for cancer therapy. The robustness of the QSAR model was validated across varying chemical profiles and trained with a dataset obtained from ChEMBL. Using this trained model, the MolPort database was used to classify active compounds. The molecular docking of the top 10 predicted inhibitors revealed that Molecule 1 and Molecule 5 exhibited the best binding affinities. Moreover, these molecules shared common binding amino acid residues with the known inhibitor CCX777, confirming that they bind at the same site. In conclusion, QSAR modeling and molecular docking successfully identified high-affinity CXCR7 inhibitors, presenting promising candidates for further experimental validation.

ACKNOWLEDGEMENT

The authors thank AIC-DSU for supporting them with the server for conducting their study.

REFERENCES

[1] F. Bray, M. Laversanne, H. Sung, J. Ferlay, R.L. Siegel, I. Soerjomataram, and A. Jemal, "Global cancer statistics 2022: GLOBOCAN estimates of incidence and mortality worldwide for 36 cancers in 185 countries," CA Cancer J Clin, vol. 74, pp. 229–263, 2024. https://doi.org/10.3322/caac.21834.

[2] M.T. Chow and A.D. Luster, "Chemokines in Cancer," Cancer Immunol Res, vol. 2, pp. 1125–1131, 2014. https://doi.org/10.1158/2326-6066.CIR-14-0160.

[3] H. Fan, W. Wang, J. Yan, L. Xiao, and L. Yang, "Prognostic significance of CXCR7 in cancer patients: a meta-analysis," Cancer Cell Int, vol. 18, 2018. https://doi.org/10.1186/s12935-018-0702-0.

[4] X. Sun, G. Cheng, M. Hao, J. Zheng, X. Zhou, J. Zhang, R.S. Taichman, K.J. Pienta, and J. Wang, "CXCL12 / CXCR4 / CXCR7 chemokine axis and cancer progression," Cancer and Metastasis Reviews, vol. 29, pp. 709–722, 2010. https://doi.org/10.1007/s10555-010-9256-x.

[5] Y. Shi, D.J. Riese, and J. Shen, "The Role of the CXCL12/CXCR4/CXCR7 Chemokine Axis in Cancer," Front Pharmacol, vol. 11, 2020. https://doi.org/10.3389/fphar.2020.574667.

[6] A. Cherkasov, E.N. Muratov, D. Fourches, A. Varnek, I.I. Baskin, M. Cronin, J. Dearden, P. Gramatica, Y.C. Martin, R. Todeschini, V. Consonni, V.E. Kuz'min, R. Cramer, R. Benigni, C. Yang, J. Rathman, L. Terfloth, J. Gasteiger, A. Richard, and A. Tropsha, "QSAR Modeling: Where Have You Been? Where Are You Going To?," J Med Chem, vol. 57, pp. 4977–5010, 2015. https://doi.org/10.1021/jm4004285.

[7] S. Forli, R. Huey, M.E. Pique, M.F. Sanner, D.S. Goodsell, and A.J. Olson, "Computational protein–ligand docking and virtual drug screening with the AutoDock suite," Nat Protoc, vol. 11, pp. 905–919, 2016. https://doi.org/10.1038/nprot.2016.051.

[8] N. Lounsbury, "Advances in CXCR7 Modulators," Pharmaceuticals, vol. 13, 2020. https://doi.org/10.3390/ph13020033.

[9] J.B. Baell, and G.A. Holloway, "New Substructure Filters for Removal of Pan Assay Interference Compounds (PAINS) from Screening Libraries and for Their Exclusion in Bioassays," J Med Chem, vol. 53, pp. 2719–2740, 2010. https://doi.org/10.1021/jm901137j.

[10] A. Nath, A. Kumer, F. Zaben, and Md.W. Khan, "Investigating the binding affinity, molecular dynamics, and ADMET properties of 2,3-dihydrobenzofuran derivatives as an inhibitor of fungi, bacteria, and virus protein," Beni Suef Univ J Basic Appl Sci, vol. 10, 2021. https://doi.org/10.1186/s43088-021-00117